\documentclass[12pt,preprint]{aastex}

\shorttitle{OT 060420}
\shortauthors{Shamir \& Nemiroff}

\def\deg{\hbox{$^\circ$}}
\def\min{\hbox{$^\prime$}}

\begin{document}

\title{OT 060420: A Seemingly Optical Transient Recorded by All-Sky Cameras}

\author{Lior Shamir}
\affil{Department of Physics, Michigan Technological University, Houghton, MI 49931}
\email{lshamir@mtu.edu}

\and

\author{Robert J. Nemiroff}
\affil{Department of Physics, Michigan Technological University, Houghton, MI 49931}

\begin{abstract}
We report on a $\sim$5th magnitude flash detected for approximately 10 minutes by two {\it CONCAM} all-sky cameras located in Cerro Pachon -- Chile and La Palma -- Spain. A third all-sky camera, located in Cerro  Paranal -- Chile did not detect the flash, and therefore the authors of this paper suggest that the flash was a series of cosmic-ray hits, meteors, or satellite glints. Another proposed hypothesis is that the flash was an astronomical transient with variable luminosity. In this paper we discuss bright optical transient detection using fish-eye all-sky monitors, analyze the apparently false-positive optical transient, and propose possible causes to false optical transient detection in all-sky cameras.
\end{abstract}
\keywords{methods: data analysis --- methods: statistical --- instrumentation: miscellaneous}

\section{Introduction}

Real-time detection of bright optical transients has an important role in modern astronomy. However, since traditional narrow-angle telescopes cover only a very small portion of the sky at a given time, one can reasonably assume that some bright short-timescale transients may not be recorded by the available narrow-angle sky surveys \citep{Nem03}.

One approach of searching for bright optical transients is by deploying and operating continuous panoramic all-sky cameras. This policy guarantees that optical transients appearing on clear nights are recorded, given that they are brighter than the limiting magnitude of the all-sky camera.

The ability of all-sky optics to observe faint sources is not comparable to even simple narrow-angle telescopes. However, since brighter optical transients are sometimes considered more interesting to science \citep{Lin87,Men87}, the approach of using fish-eye lenses covering almost the entire sky can be weighed against narrow-field robotic telescopes that are capable of recording much fainter sources, but cover only a narrow field of view.

Searching and recording bright astronomical transients is one of the main purposes of the deployment of the {\it Night Sky Live} network \citep{Nem99}, which consists of 11 nodes called {\it CONCAM}, located in some of the world's premier observatories and covering almost the entire global night sky. Each CONCAM node incorporates an SBIG ST-8 or ST-1001E CCD camera, a Nikon FC-E8 or SIGMA F4-EX 8mm fish-eye lens and an industrial PC. One 180-second exposure is taken every 236 seconds, and the resulting 1024$\times$1024 FITS images are transmitted to the main server where they are copied to the public domain and can be accessed at {\it http://nightskylive.net}. The images are processed by a software that finds and alerts on optical transients \citep{Sha06}. The limiting stellar magnitude is dependent on the camera, the lens and the site, and can get to 6.8 near the image center \citep{Sha05a}. 

On April 20th at around 00:20 UT, a bright transient candidate was recorded simultaneously at Cerro Pachon -- Chile, and Canary Islands -- Spain \citep{Nem06a}. The transient candidate seemed to have a point spread function, was recorded in the same location by two distant cameras, and seemed to rotate with the sky for three consecutive frames. This unusual observation indicated that the flash seemed at first to be a true bright optical transient \citep{Nem06a}. However, a second report indicated that the transient was not recorded by a third camera \citep{Sme06}, and with the somewhat shorter point spread function of the transient candidate led to the hypothesis that the recorded source was probably not astronomical \citep{Nem06b}.

In this paper we examine the recorded source and describe proposed causes of such an unusual event that seemed like a true astronomical transient. We also discuss the efficacy and limitations of all-sky panoramic search for bright optical transient using fish-eye lenses. In Section~\ref{transient_detection} we describe the all-sky transient detection mechanism, in Section~\ref{060420} we describe the transient candidate, and in Section~\ref{nature} we propose theories for the cause of this seemingly convincing observation.

\section{Transient Detection Using All-sky Cameras}
\label{transient_detection}

The detection of optical transients in all-sky images requires several logical steps.  The first is building a database of canonical images that can be compared to a given image. This is done by co-adding several images taken on a clear night at the same sidereal time. A detailed description of this mechanism is described in \citep{Sha05b}.

The next step is the rejection of pixels dominated by cosmic-ray generated counts. Rejecting cosmic rays by comparing two images of the same field \citep{Sha92,Fix00} cannot be used for this purpose since this method inherently rejects flashes appearing in just one image, and might lead to the rejection of some true transients. Since the use of this approach is inappropriate for the described system, rejections should make use of the non-point source nature of cosmic ray splashes. We chose to use a fuzzy logic-based algorithm for cosmic ray hit rejection from single images \citep{Sha05e}. This algorithm is reasonably accurate, and provides low computational complexity allowing it to process images in a relatively short time.

Next, bright planets are also rejected using a star recognition algorithm designed to find astronomical objects in wide angle frames \citep{Sha05d}. In the same fashion, high-amplitude variable stars (dM$>$0.7) are also rejected from the image.

After cosmic-ray hits and bright planets are rejected, the system searches for PSFs that are a given $\sigma$ brighter than the local background (for the Night Sky Live transient detection mechanism we use a threshold of 40$\sigma$). Each PSF is then compared with the canonical frame. If the PSF does not appear in the canonical image, it is assumed to be a potential transient candidate.

Once a flash is recorded, the image is compared to the image taken by another camera covering the same area of the sky where the transient was detected. It is also compared to previous images to check if the flash is persistent and rotates with the sky. If a flash appears to be rotating with the sky for several consecutive images, and/or recorded simultaneously by two distant cameras, the flash is alerted as an optical transient candidate. A more detailed description of the all-sky transient detection mechanism is available at \citep{Sha06}.

\section{Optical Transient 060420}
\label{060420}

\begin{figure}
\includegraphics{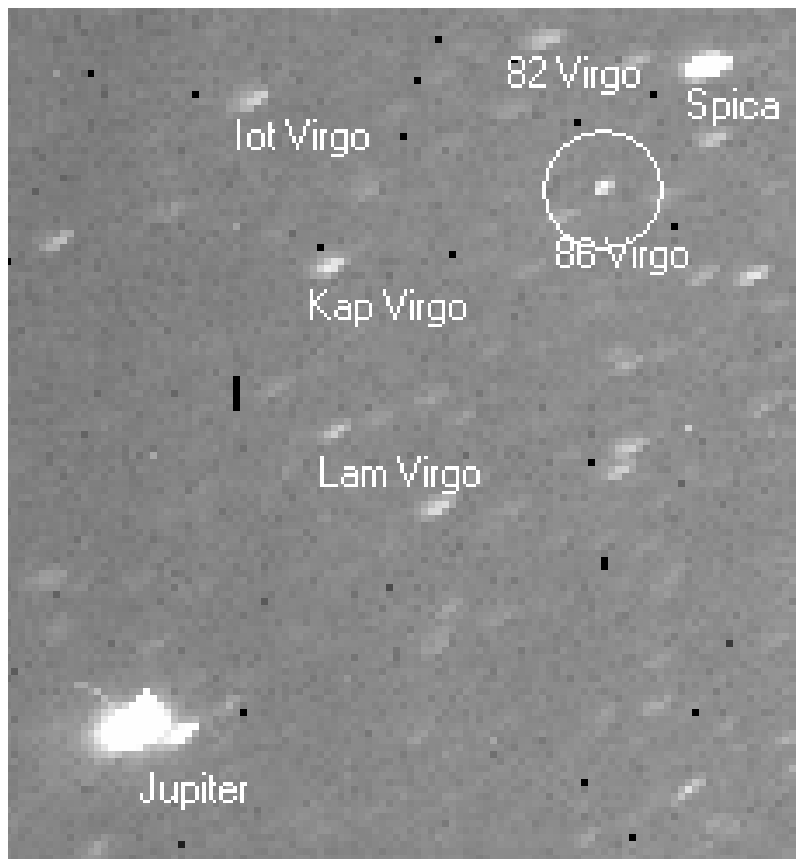}
\caption{Image recorded at Cerro Pachon on 4/20/06 at 00:19:43 UT (180 second exposure).}
\label{cp1}
\end{figure}

\begin{figure}
\includegraphics{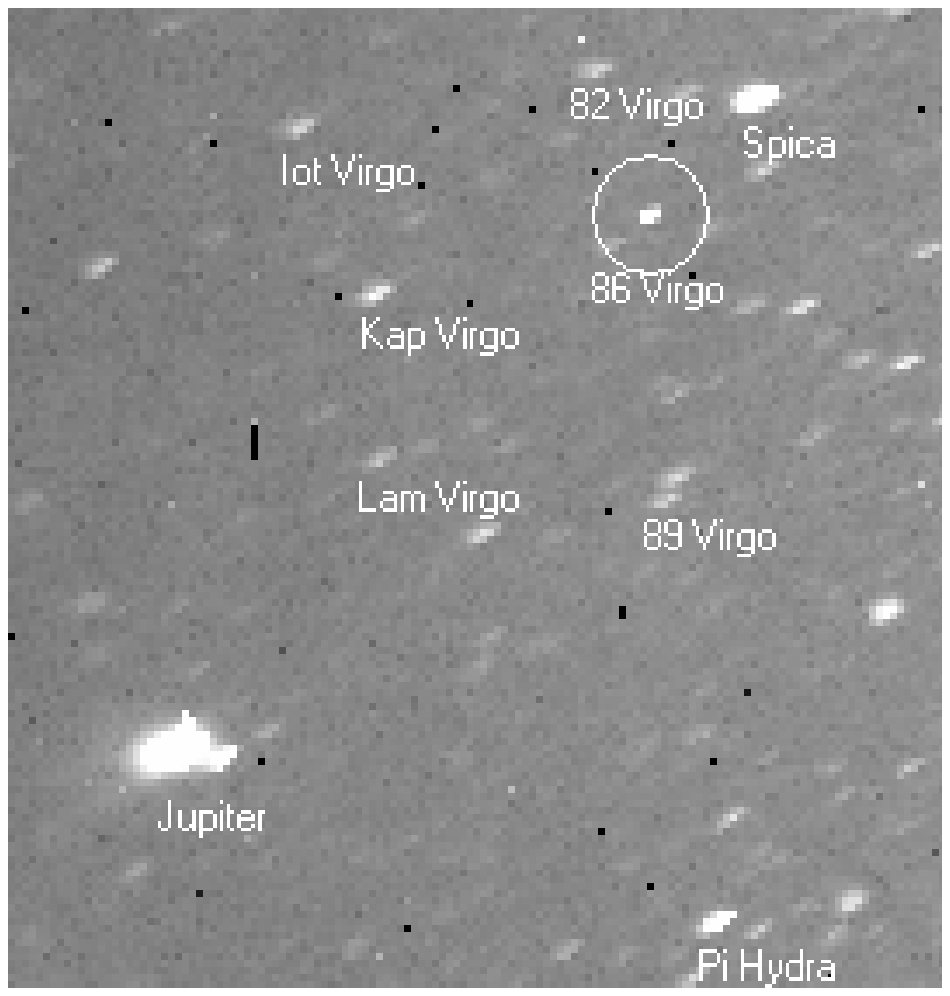}
\caption{Image recorded at Cerro Pachon on 4/20/06 at 00:23:39 UT (180 second exposure)}
\label{cp2}
\end{figure}

\begin{figure}
\includegraphics{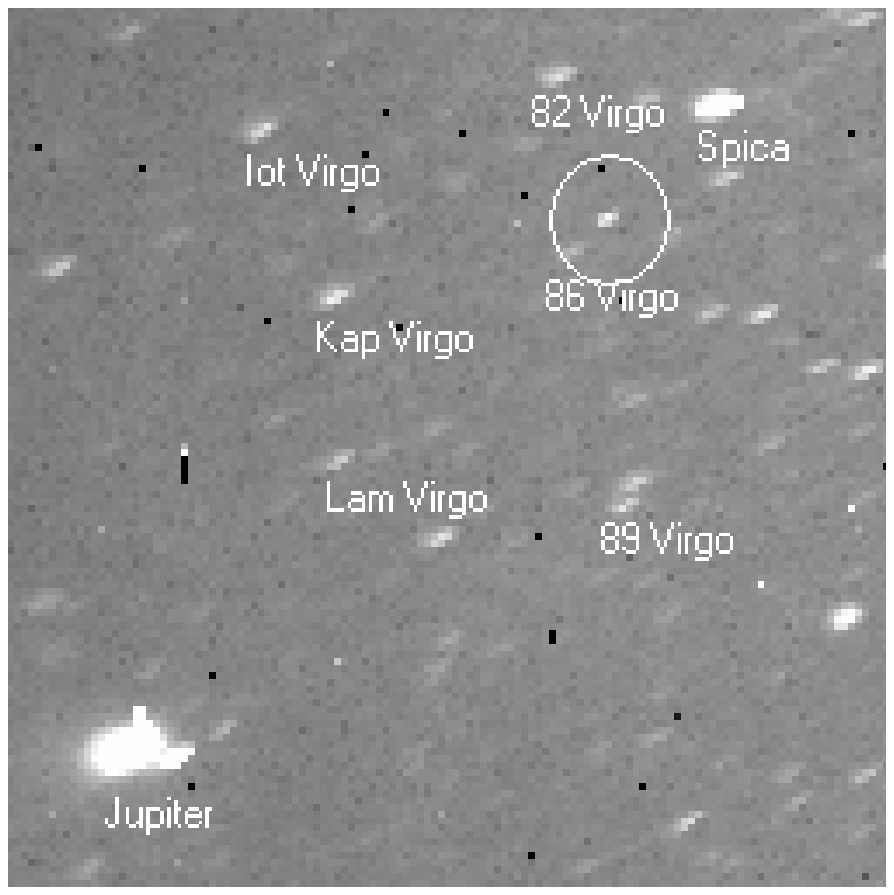}
\caption{Image recorded at Cerro Pachon on 4/20/06 at 00:27:35 UT (180 second exposure)}
\label{cp3}
\end{figure}

On the night of 4/20/2006, a flash was recorded by two CONCAM all-sky cameras located in Cerro Pachon, Chile and La Palma, Spain. The flash was detected by the transient detection software described in \citep{Sha06}. The first image recording the flash was taken in Cerro Pachon at 00:19:43 UT, and the flash was clearly recorded by the next two images taken in Cerro Pachon at 00:23:39 and 00:27:35. The series of the three images showed that the flash maintained its position comparing to nearby stars, and seemed to be rotating with the sky. The location of the flash was around RA: 13h 40$\min$, DEC: -11$\deg$ 40$\min$, with accuracy of $\pm$10$\min$. The point spread function of the flash seemed to ``trail'' to the same direction as its neighboring stars. Figures~\ref{cp1},~\ref{cp2} and~\ref{cp3} show the flash recorded in Cerro Pachon.

\begin{figure}
\includegraphics{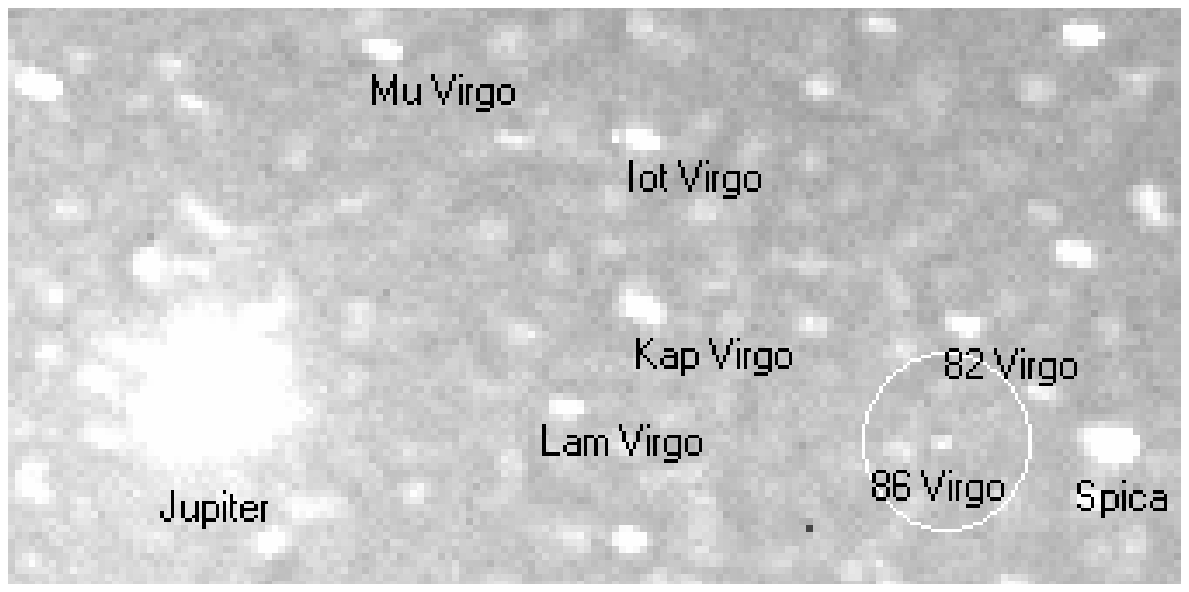}
\caption{Image recorded at La Palma on 4/20/06 at 00:19:43 UT (180 second exposure).}
\label{ci1}
\end{figure}

\begin{figure}
\includegraphics{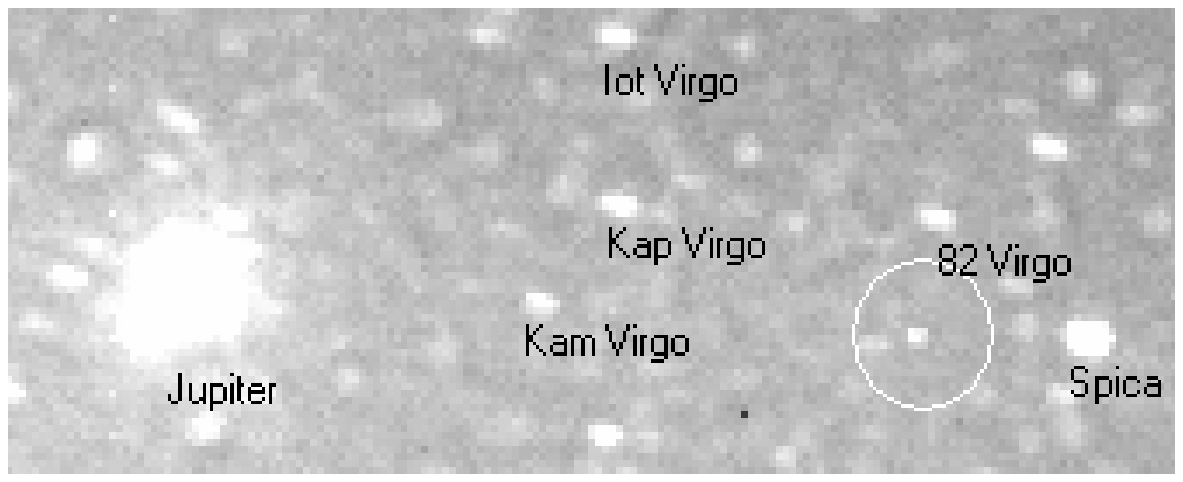}
\caption{Image recorded at La Palma on 4/20/06 at 00:23:41 UT (180 second exposure)}
\label{ci2}
\end{figure}

Two images taken at the same time in La Palma also showed a flash recorded at the exact same location. The two images were taken at 00:19:43 UT and 00:23:41 UT. Figures~\ref{ci1} and~\ref{ci2} show the flash recorded in La Palma. The quality of these images is marginal due to the presence of moisture on the lens, but a bright source at the same location is still noticeable.

A third image (that should have been taken at 00:27:35) was not taken in La Palma due to bandwidth constraints at the site. Since all CONCAM stations are synchronized, an exposure is taken by all active stations at the exact same time. However, if the transmission of the previous image file from a certain station to the main server has not been completed, an image is not taken by that station, and the next exposure is started at the next 236-second interval. The idea of this policy is to avoid a cumulative delay in cases where the average transmission time of a single image is longer than 236 seconds. For instance, if the transmission of an image from a certain CONCAM station to the main server takes 6 minutes, after 8 hours of operation images will arrive at the server $\sim$4 hours after they were taken. Since the CONCAM stations are often used as cloud monitors \citep{Sha05b}, it is important that images will be available in nearly real time. Therefore, stations with a slow internet connection do not always transmit all images to the server. Since the internet connection in La Palma is slow, some images are not transmitted to the main server, and are removed from the local disk to free some disk space for newer images.

According to the images, the recorded flash was dimmer in the first and last image, and was brighter in the second image (taken at 00:23:39). This is consistent in the images taken by both CONCAM stations (except from the last image in La Palma, which was not taken). The approximate magnitude of the flash can be deduced by comparing it to known nearby stars. This analysis, however, is dependent on the color of the flash, which is unknown. Table~\ref{magnitude} shows the approximate peak magnitude of the recorded flash assuming its color was A, F, G, K or M.

Comparison of the location of the flash in the images taken in Cerro Pachon to its location in the images taken in La Palma shows no parallax. Since the pixel size of a Night Sky Live image is 10$\min$ \citep{Sha05a}, and the distance between La Palma and Cerro Pachon is $\sim$10000 kilometers, sub-pixel positioning cannot lead to distance of less than $3.4\cdot10^{6}$ km.

\begin{figure}
\includegraphics{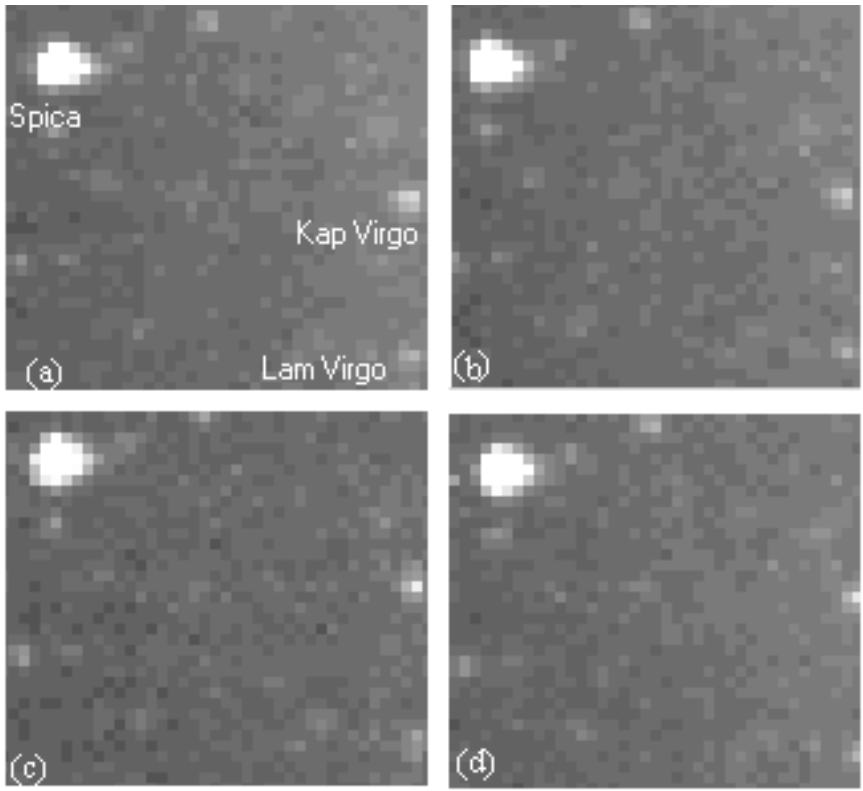}
\caption{Image recorded at Paranal on 4/20/06 at 18:44 (a), 21:38 (b), 24:27 (c), and 28:54 (d).}
\label{paranal}
\end{figure}

Another all-sky MASCOT \citep{Mas06} camera located in Paranal Observatory was also operating at the same time, but did not record the flash \citep{Sme06}. Figure~\ref{paranal} shows the same field recorded by the MASCOT camera at the same time the flash was recorded by CONCAM.

\begin{figure}
\includegraphics{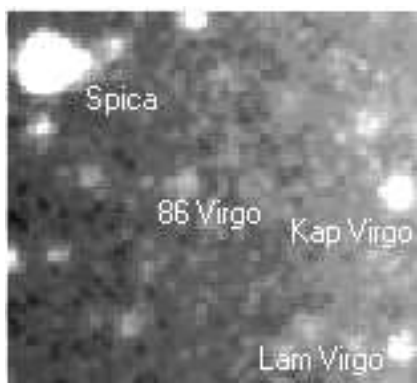}
\caption{The co-added image of the four images shown in Figure~\ref{paranal}}
\label{co_added}
\end{figure}

Unlike the 180-second exposure of CONCAM, MASCOT takes a 90-second exposure image every $\sim$3 minutes, and the limiting magnitude is much brighter than that of CONCAM. As can be seen from the images, the nearby 5.5th magnitude 86 Virgo, that is seen clearly in CONCAM images, does not appear in the MASCOT images. In order to search for the flash in the MASCOT images, the images were co-added to improve the signal-to-noise ratio. The resulting co-added image is shown in figure~\ref{co_added}.

As can be seen in Figure~\ref{co_added}, 86 Virgo can be clearly seen near the center of the co-added image, while the flash recorded by CONCAM images does not appear. Therefore, the authors of this paper suggest that the recorded flash is not related to an astronomical source \citep{Nem06b}.

An attempt to track a possible source in the approximate location of the flash did not provide a detection of any unusual source \citep{Pol06}. The search was performed three days after the flash was recorded using an 18-inches telescope located at Wise observatory, Israel. 

\section{The Nature of the Recorded Flash}
\label{nature}

In this section we propose several theories concerning the cause of the flash, in an attempt to understand the nature of this seemingly convincing false positive detection.

\subsection{Cosmic Ray Hit}

One explanation of the flash recorded by the three images in Cerro Pachon is that it is simply a coincidental series of cosmic-ray hits, such that the three cosmic-ray hits seem like one flash rotating with the sky.  Coincidental cosmic-ray hits have been previously reported as a potential cause of false positives in cases where a large number of astronomical images are processed, so that seemingly astonishing discoveries \citep{Sah01} are sometimes turn out to be nothing but a coincidental series of cosmic-ray hits \citep{Sah02}. Due to the possibility of false positives due to cosmic ray hits, sky surveys searching for supernovae usually record several exposures of the same field before triggering an alert of a potential discovery \citep{Bal06,Kim02}.

A Night Sky Live image contains, in average, 6 cosmic ray hits brighter than 20$\sigma$ inside the field of view \citep{Sha05e}. Since the field of view contains $\sim$600,000 pixels, the probability that a cosmic-ray hit will appear at a certain pixel is $\sim10^{-5}$. Therefore, the probability that three consecutive bright cosmic ray hits will appear at the exact same pixel in three consecutive frames is $\sim 6 \cdot (10^{{-5})^2}=6 \cdot 10^{-10}$. 

An individual fish-eye CONCAM camera records $\sim$140 images per night. However, since only $\sim25\%$ of the image are clear \citep{Per05}, an average of 35 images per night can be used. Therefore, the probability of recording a flash that appears in three consecutive images at the same geocentric coordinates is $\sim2\cdot10^{-8}$ for a single night, and $\sim 7.6 \cdot 10^{-6}$ for one year. Since the software has been active for three years, the probability to record one such flash is $\sim2.3\cdot10^{-5}$. Since the software is running in two CONCAM stations (Cerro Pachon and La Palma), the probability that such a flash will be recorded in three years by any of the stations is $\sim4.6\cdot10^{-5}$, which is approximately 3.9$\sigma$.

If we also include the 2 marginal-quality images taken in La Palma, there are 5 images in which the flash was recorded in the exact same location in the sky. Based on the analysis above, the probability that cosmic-ray hits will appear at the same location in five images is $\sim 6 \cdot (10^{{-5})^4}=6 \cdot 10^{-20}$. The probability that such an event will be recorded in three years is therefore $\sim 4.6 \cdot 10^{-15}$, or $\sim 7.7\sigma$.


Assuming inaccuracy of one pixel (meaning that the cosmic ray hit can lay anywhere in the 9$\times9$ grid around the exact position), the probability that two cosmic-ray hits will appear at the same location in two consecutive images is $\frac{9}{600000}\cdot6\simeq10^{-4}$. Therefore, the probability that three consecutive bright cosmic ray hits will appear at the same geocentric coordinates in three consecutive frames is $\sim 6 \cdot (10^{{-4})^2}=6 \cdot 10^{-8}$. With 2 all sky cameras operating simultaneously for three years, the probability to record such a flash is $\sim1.5\cdot10^{-3}$, which is approximately 3$\sigma$.


By including also the 2 marginal-quality images taken in La Palma, the probability that cosmic-ray hits will appear approximately (one pixel accuracy) at the same location in five images is $\sim 6 \cdot (10^{{-4})^4}=6 \cdot 10^{-16}$, and the probability to record such an event after three years of operation is $\sim 4.6 \cdot 10^{-11}$, which is approximately 6.5$\sigma$.

\subsection{A Glint From an Artificial Object}

Another possible explanation to the flash is light reflected by satellites or space junks. Glints from satellites and other artificial objects are recorded frequently by CONCAM fish-eye cameras \citep{Sha05c,Nem05}. In many cases, especially in the case of geosynchronous satellites, the glint can be recorded in more than one image, and can sometimes persist in 5 consecutive images \citep{Sha05c}.

What makes the flash discussed in this paper different from most satellite glints is that the flash seemed to rotate with the sky. However, some previously recorded flashes coming from the same location in the sky \citep{Hal87}, and therefore suspected to be astronomical sources, appeared later to be caused by artificial objects \citep{Sch87}.

Since artificial satellites have a relatively small orbit, a satellite glint recorded at one location (such as Cerro Pachon) is not expected to be recorded also at a distant location (such as La Palma). However, the quality of the images recorded in La Palma is marginal, and moisture on the primary lens often causes effects that look very much like astronomical sources.

Comparison with the database provided by {\it Heavens-Above} did not provide expected satellite glints at the location of the flash. However, previous satellite glints recorded by CONCAM were also not included in {\it Heavens-Above} database \citep{Sha05c}, and therefore the possibility that the flash was caused by a satellite glint cannot be rejected based on this criterion.

\subsection{A Variable Source}

Another explanation to the flash is that the recorded source was indeed of astronomical nature, but was not recorded by the MASCOT camera. This can be caused by an optical transient with a variable magnitude. MASCOT camera is not synchronized with the Night Sky Live network, and therefore the exposures taken by MASCOT are taken at different times. CONCAM cameras, however, are synchronized so that all cameras take an exposure at the same time. The times of the exposures taken by CONCAM (at Cerro Pachon and La Palma) and MASCOT are given in Table~\ref{CONCAM} and~\ref{MASCOT} respectively.

If the magnitude of the flash was not constant, it could have been recorded by CONCAM exposures without being recorded by MASCOT. For instance, consider the following scenario described in Table~\ref{scenario}.

The scenario described in Table~\ref{scenario} assumes that the flash recorded by CONCAM was not constant, but a series of several shorter flashes. According to that scenario, the flash was expected to be recorded by the two CONCAM cameras, but not by the MASCOT camera since it was not present (or was too dim) while a MASCOT exposure was taken.

As explained in Section~\ref{060420}, the limiting magnitude of the MASCOT camera is significantly brighter than CONCAM, and the luminosity of the flash is just around its limit. Therefore, even if the flash was partially covered by a MASCOT exposure, it is reasonable that it will not be noticeable in the resulting images. This explanation also supports the observation described in Section~\ref{060420}, according which the PSFs of the flash are shorter than the PSFs of nearby stars.

\subsection{Meteors}

An event of several flashes appearing around the same location can also be a result of a minor meteor shower. Meteors are often recorded by CONCAM \citep{Raf01} or other all-sky cameras \citep{Spu04}. When the trajectory of the meteor is oriented in the direction of the camera, the meteor leaves a point spread function very similar to stars in the image \citep{Bro02}. The meteor showers active on the night of April 20th are Piscids \citep{Mcb03a}, Lyrids \citep{Art97}, and Aquarids \citep{Mcb03b}. Since non-sporadic meteors are coming from the same area in the sky (the radiant), it might happen that three different meteors appear in three consecutive exposures at around the same location. This is more likely to happen during a meteor shower with a high meteor rate. On the night of April 20, however, no unusual meteor activity was recorded. On nights that are not peaks of meteor showers, about 10 noticeable point meteors are recorded every year. Therefore, the probability that three sporadic point meteors will be recorded in three consecutive images at the same location is negligible, and equals to $\sim5.5\cdot10^{-16}$.

\section{Conclusion}

We analyzed a flash that was recorded by an all-sky camera located in Cerro Pachon, and seemed to rotate with the sky in three images. This observation was supported by images of marginal quality provided by another all-sky camera located in La Palma, but was not recorded by a third camera located in Paranal. Although the limiting magnitude of the Paranal camera is significantly brighter than the other cameras, co-adding the images did not introduce any sign of an optical source.

Since the probability that a rare match of three cosmic-ray hits is somewhat low, we believe that the most likely explanation is that the flash was caused by satellites or other artificial objects.

We therefore suggest that artificial non-astronomical sources might be recorded by astronomical all-sky cameras in a fashion that makes them seem like true optical transients. We consider this as a concern when using all-sky cameras for the purpose of bright optical transient detection, and for the evaluation of the scientific nature of future transient events recorded by such instruments.

\clearpage

\clearpage

\clearpage

\clearpage

\clearpage

\clearpage

\clearpage

\clearpage

\begin{table}
\begin{center}
\begin{tabular}{lc}
\hline  Color & Magnitude \\
\hline 
A & 4.3 $\pm$0.2 \\
F & 4.5 $\pm$0.2 \\
G & 4.7 $\pm$0.2 \\
K &  4.9 $\pm$0.3 \\
M &  5.2 $\pm$0.3 \\
\hline
\end{tabular}
\caption{The peak magnitude of the flash}
\label{magnitude}
\end{center}
\end{table}

\clearpage

\begin{table}
\begin{center}
\begin{tabular}{lcc}
\hline  
Exposure No. & Start (UT) & End (UT) \\
\hline 
1 & 00:19:43 & 00:22:43 \\
2 & 00:23:39 & 00:26:39 \\
3 & 00:27:36 & 00:30:36 \\
\hline
\end{tabular}
\caption{Start and end of the exposures taken by CONCAM on 4/20/2006}
\label{CONCAM}
\end{center}
\end{table}

\clearpage

\begin{table}
\begin{center}
\begin{tabular}{lcc}
\hline  
Exposure No. & Start (UT) & End (UT) \\
\hline 
1 & 00:18:44 & 00:20:16 \\
2 & 00:21:38 & 00:23:08 \\
3 & 00:24:27 & 00:25:57 \\
4 & 00:27:24 & 00:28:54 \\
\hline
\end{tabular}
\caption{Start and end of the exposures taken by MASCOT on 4/20/2006}
\label{MASCOT}
\end{center}
\end{table}

\clearpage

\begin{table}
\begin{center}
\begin{tabular}{lcc}
\hline  
Flash No. & Start (UT) & End (UT) \\
\hline 
1 & 00:20:16 & 00:21:36 \\
2 & 00:23:10 & 00:24:30 \\
3 & 00:29:00 & 00:30:20 \\
\hline
\end{tabular}
\caption{A possible scenario of several flashes coming from the same location}
\label{scenario}
\end{center}
\end{table}

\end{document}